\documentclass[twocolumn,prl,superscriptaddress]{revtex4-2}

\usepackage{graphics}
\usepackage{color}

\usepackage{amssymb}
\usepackage{amsmath}
\usepackage{overpic}
\usepackage{epstopdf}

\usepackage{bm}
\usepackage{graphicx}

\usepackage{color}
\usepackage[normalem]{ulem}

\newcommand{\tpt}{Ta$_{4}$Pd$_{3}$Te$_{16}$}

\begin{document}

\title{
Incommensurate two-dimensional checkerboard charge density wave in the low dimensional superconductor \tpt
}

\author{Zhenzhong Shi}
\affiliation{Department of Physics, Duke University, Durham, NC 27708, USA}
\author{S. J. Kuhn}
\affiliation{Department of Physics, Duke University, Durham, NC 27708, USA}
\author{F. Flicker}
\affiliation{Rudolph Peierls Centre for Theoretical Physics, University of Oxford, Department of Physics, Clarendon Laboratory, Parks Road, Oxford OX1 3PU, UK}
\author{T. Helm}
\affiliation{Max Planck Institute for Chemical Physics of Solids, Dresden, Germany}
\author{J. Lee}
\affiliation{Cornell High Energy Synchrotron Source, Cornell University, Ithaca, NY 14853, USA}
\author{William Steinhardt}
\affiliation{Department of Physics, Duke University, Durham, NC 27708, USA}
\author{Sachith Dissanayake}
\affiliation{Department of Physics, Duke University, Durham, NC 27708, USA}
\author{D. Graf}
\affiliation{National High Magnetic Field Laboratory, Florida State University, Tallahassee, FL 32310, USA}
\author{J. Ruff}
\affiliation{Cornell High Energy Synchrotron Source, Cornell University, Ithaca, NY 14853, USA}
\author{G. Fabbris}
\affiliation{Advanced Photon Source, Argonne National Laboratory, Argonne, IL 60439, USA}
\author{D. Haskel}
\affiliation{Advanced Photon Source, Argonne National Laboratory, Argonne, IL 60439, USA}
\author{S. Haravifard}
\affiliation{Department of Physics, Duke University, Durham, NC 27708, USA}
\affiliation{Department of Mechanical Engineering and Materials Science, Duke University, Durham, North Carolina 27708, USA}

\date{\today}

\begin{abstract}

We report the observation of a two-dimensional (2D) checkerboard charge density wave (CDW) in the low-dimensional superconductor \tpt. By determining its CDW properties across the temperature-pressure ($T$-$P$) phase diagram and comparing with prototypical CDW materials, we conclude that \tpt\ features: a) an incommensurate CDW with a mixed character of dimensions (Q1D considering its needle-like shape along the \bm{$ \mathrm{b} $}-axis, Q2D as the CDW has checkerboard wavevectors, and 3D because of CDW projections along all three axes); and b) one of the weakest CDWs compared to its superconductivity (SC), i.e. enhanced SC with respect to CDW, suggesting an interesting interplay of the two orders. 

\end{abstract}

\maketitle

Charge density waves (CDWs), electronic instabilities originally considered in one dimension (1D) \cite{Peierls1955}, have now been found in various forms and exist not only in quasi-one dimensional (Q1D) materials but also in quasi-two dimensional (Q2D) and three dimensional (3D) systems \cite{Monceau2012,Chen2016}. Q1D systems such as the transition metal trichalcogenide compounds are generally expected to be unstable to CDW formation via the Peierls instability, i.e. Fermi surface (FS) nesting (FSN) \cite{Gruner1988}. However, in systems with higher dimensions, such as the transition metal dichalcogenides (Q2D) and cubic system R$_3$T$_4$X$_{13}$ (3D) \cite{Klintberg2012,Goh2011}, FSN is not the natural explanation for CDW formation. A single wavevector will always nest the two points of the FS onto one another in 1D; in 2D and 3D a special shape of FS would be required to match large portions of the FS onto itself. Instead, CDWs in higher dimensional materials such as NbSe$_{2}$, often have a more complex origin involving the momentum and orbital dependence of the electron-phonon coupling matrix \cite{Johannes2008,Zhu2015,Flicker2015}. 

The electron-phonon coupling required for the CDW order may also facilitate superconductivity (SC) based on the standard Bardeen Cooper Schrieffer (BCS) mechanism \cite{Bardeen1957a}, and the two orders are often found to be proximal or to coexist \cite{Wu2011,Wu2015,Comin2016,Moncton1975,Monceau1976,Monceau2012,Chen2016}. However, the interplay of the CDW and SC may extend well beyond the BCS picture, where the pairing mechanism may not be phonon mediated and the SC becomes unconventional. In cuprate, for example, the relative strength of the SC with respect to the charge order is much stronger than that in conventional superconductors \cite{Comin2016} and the interplay of charge order and SC is still highly debated \cite{Gabovich2002,Norman2011,Fradkin2015,Shi2020,Shi2020a}. It is possible that SC in some unconventional superconductors could be enhanced by quantum fluctuations accompanied by the suppression of the CDW to a quantum critical point (QCP) \cite{Gruner2017,Lee2020}. In reality, however, it is extremely difficult to discern the different yet possibly intertwined mechanisms contributing to the interplay of the CDW and SC, including dimensionality and critical fluctuations. Here we report a CDW system that may serve as an ideal platform for such studies.

Featuring three chains of PdTe$_{2}$, TaTe$_{3}$, and Ta$_{2}$Te$_{4}$ along its \bm{$ \mathrm{b} $}-axis, aligned into planes perpendicular to \bm{$ \mathrm{c} $}, the monoclinic crystalline structure of \tpt\ with lattice parameters \textit{a} = 17.687 \AA, \textit{b} = 3.735 \AA, \textit{c} = 19.510 \AA, and $ \beta $ = 110.42$^{\circ}$ \cite{Jiao2014} is reminiscent of those of trichalcogenide compounds such as ZrSe$_{3}$, TaSe$_{3}$, NbSe$_{3}$, and TaS$_{3}$. However, transport measurements on \tpt\ indicate resistivity ratios along the \bm{$ \mathrm{a^{*}} $}:\bm{$ \mathrm{b} $}:\bm{$ \mathrm{c} $} directions of $4:1:13$ at room temperature ($T$), evolving to $10:1:20$ just above the superconducting transition $T_{c}$ $\sim$ 4.6 K \cite{Helm2017}. The ratios suggest a rather weak anisotropy, and the material itself is better thought of as having a character of mixed dimensionality: Q1D (along \bm{$ \mathrm{b} $}), Q2D (cleavage plane perpendicular to \bm{$ \mathrm{c} $}), and 3D (relatively weak anisotropy), which, as we will show below, underlies an interesting 2D checkerboard CDW in this system with wavevectors projected along all three axes.

Previous studies of \tpt\ have revealed an anisotropic SC ground state \cite{Du2015,Jiao2016}. Under pressure ($P$), a dome-shaped SC phase with maximum $T_{c}$ $\sim$ 6 K at 0.2 $\sim$ 0.3 GPa was discovered \cite{Pan2015,Jo2017}. Some evidence suggests that the SC order parameter in \tpt\ might have a $d$-wave symmetry and thus an unconventional origin \cite{Du2015,Pan2015,Pang2018}. Others have argued that the experimental findings may be explained by the more conventional $s$-wave SC order parameter and the multi-band nature of the compound \cite{Fan2015,Li2016,Singh2014}. While phase-sensitive measurements are needed to eventually reveal the pairing symmetry, the SC in \tpt\ is also interesting because of hints of its proximity to a CDW phase \cite{Fan2015,Li2016,Singh2014,Helm2017,Chen2015}. NMR/NQR \cite{Li2016} and thermodynamic measurements \cite{Helm2017} suggest a CDW transition temperature ($T_{\textrm{CDW}}$) of $\sim$ 20 K, while Raman scattering measurements support the emergence of CDW fluctuations below 140 K $\sim$ 200 K \cite{Chen2015}. Additionally, STM measurements have shown period-4 commensurate CDW-like stripes along the \bm{$ \mathrm{b} $}-axis that coexist with SC \cite{Fan2015}. However, the suggested wavevector is seemingly at odds with transport measurements, which reveal a resistivity anomaly primarily along the \bm{$ \mathrm{a^{*}} $}-direction while being much weaker along the \bm{$ \mathrm{b} $} and \bm{$ \mathrm{c} $} directions \cite{Helm2017}. Therefore, a direct measurement of the bulk CDW in \tpt\ is critical for reconciling these experimental results and understanding the interplay of SC and CDW in this material. 
 
In this paper, we report the direct observation of a delicate CDW in \tpt\ using synchrotron X-ray diffraction (See Supplemental Material \cite{Supplemental}). We establish the $T$-$P$ phase diagram, which reveals a low $T_{\textrm{CDW}}$ $\sim$ 16 K at ambient $P$, quickly suppressed with increasing $P$, becoming undetectable where $T_{c}$ reaches its maximum. Our de Haas-van Alphen (dHvA) oscillation measurements (See Supplemental Material \cite{Supplemental}) do not show any sign of FS reconstruction through the CDW transition, consistent with an earlier report \cite{Helm2017}. The observed CDW has a checkerboard pattern in real space, which resembles more closely the checkerboard CDW in quasi-2D systems such as NbSe$_{2}$ \cite{Chen2016} rather than those in typical Q1D systems such as NbSe$_{3}$, where multiple CDWs, if they exist, are typically unidirectional and independent \cite{Monceau2012}. However, unlike the CDWs seen in most Q2D systems, the CDW in \tpt\ does not lie within the Q2D planes. These peculiar dimensional properties may explain its weak CDW, as we discuss in detail below. 

%
\begin{figure}
\centering
\includegraphics[width=8.5cm]{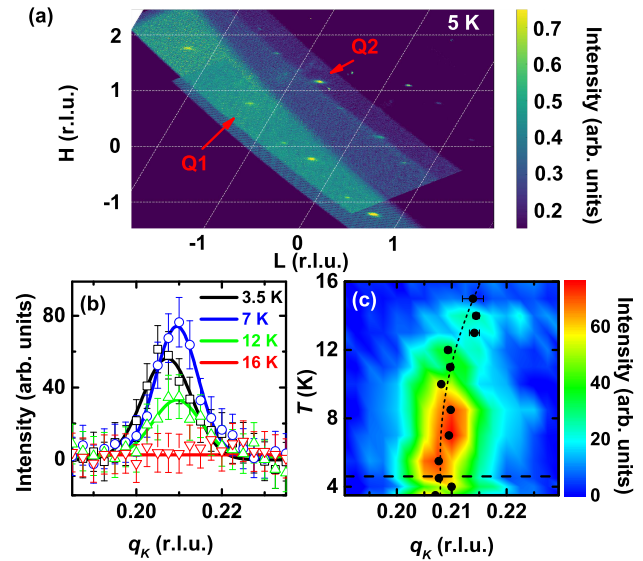}
\caption{(a) 2D diffraction pattern along \bm{$ \mathrm{H} $} and \bm{$ \mathrm{L} $} at $K$=4.21 for $T$=5 K, where $H$, $K$, and $L$ positions are given in reciprocal lattice units (r.l.u.). Two CDW peaks, labeled as \bm{$ \mathrm{Q1} $} and \bm{$ \mathrm{Q2} $}, appear in all the Brillouin zones covered in the experiment: \bm{$ \mathrm{Q1} $} $ \sim $ (-0.20, 0.21, -0.30) and \bm{$ \mathrm{Q2} $} $ \sim $ (0.20, 0.21, 0.30). Same observation is made at $ K = 2.21 $. (b) The CDW peak [(-1, 4, 1) + (-0.20, $q_{k}$, -0.30)] \bm{$ \mathrm{Q1} $} along \bm{$ \mathrm{K} $} for several $T$, with the 30 K background trace subtracted. Solid lines are Lorentzian fits to data. Error bars correspond to 1 SD (square root of number of counts). (c) Contour plot of the background-subtracted \bm{$ \mathrm{Q1} $} CDW intensity along \bm{$ \mathrm{K} $}. Black data points indicate the peak centers of Lorentzian fits at each $T$. The short-dashed line guides the eye. The horizontal dashed line indicates $ T_{c} $ = 4.6 K \cite{Jiao2014}. 
\label{Fig2}}
\end{figure}

We obtained the most direct evidence of a checkerboard CDW in \tpt\ using ambient-$P$ X-ray diffraction \cite{Supplemental}, which provides an unambiguous identification of the bulk CDW and its wavevectors. Using a high-dynamic-range area detector, we scanned wide areas across multiple Brillouin zones in many directions, including along the wavevector [0 0.25 0] suggested by a previous STM study \cite{Fan2015}. However, our X-ray scans did not reveal any CDW peak along this direction. Instead, we identified two peaks at wavevectors \bm{$ \mathrm{Q1} $} = [-0.2 0.21 -0.3] and \bm{$ \mathrm{Q2} $} = [0.2 0.21 0.3] in all the Brillouin zones within our detection range [Fig. \ref{Fig2}(a)]. The integrated intensity of the CDW peak is $\sim 10^{-3} $ the strength of that at the nearby Bragg peak. The \bm{$ \mathrm{Q1} $} and \bm{$ \mathrm{Q2} $} wavevectors are related by a $180^\circ$ rotation about the \bm{$ \mathrm{b} $}-axis. Together they constitute a commensurate pattern of the superlattice along the \bm{$ \mathrm{H} $} and \bm{$ \mathrm{L} $} (or \bm{$ \mathrm{a^{*}} $} and \bm{$ \mathrm{c^{*}}$}) directions. 

We repeated the measurements at different $T$ from 3.5 K up to 30 K. Figure \ref{Fig2}(b) shows the line-cuts and their Lorentzian fits along the \bm{$ \mathrm{K} $} direction for the \bm{$ \mathrm{Q1} $} CDW near the (-1 4 1) Bragg peak for a few selected $T$. The entire set of data consisting of the line-cuts at all measured $T$ is summarized in a color contour plot in Fig.~\ref{Fig2}(c). It is clear that, with $T$ increasing from 3.5 K, the \bm{$ \mathrm{Q1} $} CDW peak intensity first increases and plateaus at 5 K $\sim$ 9 K, then diminishes and becomes undetectable at $ T_{\textrm{CDW}}$ $\sim$ 16 K, and this is accompanied by a slight shift in the peak position. Our results are consistent with the $ T_{\textrm{CDW}}$ $\sim$ 20 K inferred from NMR/NQR \cite{Li2016} and transport measurements \cite{Helm2017}. We do not see any evidence of a static CDW above $\sim$ 16 K, though our measurements cannot rule out the possibility of fluctuating CDW order at higher $T$. The signatures of phonon anomalies at higher $T$ \cite{Chen2015} therefore might be attributed to a pseudogap phase, similar to that in NbSe$_{2}$ \cite{Borisenko2008}. A comparison between $T_{\textrm{CDW}}$ ($ \sim $16 K) and the energy gap ($ \Delta $ = 20 $\sim$ 30 meV) observed by STM \cite{Fan2015} suggests that the CDW in \tpt\ is in the strong-coupling limit (3.52$ k_{B}T_{c} \ll 2\Delta$), again similar to the case in NbSe$ _{2} $ \cite{Flicker2015} and cuprates \cite{Comin2016}. It is interesting to point out that our observation is consistent with an early band structure study on rare-earth tellurides which suggests that a checkerboard state would occur with sufficiently low $T_{\textrm{CDW}}$, while a unidirectional (stripe) CDW is favored with higher $T_{\textrm{CDW}}$ \cite{Yao2006}.

%
\begin{figure}
\centering
\includegraphics[width=7.0cm]{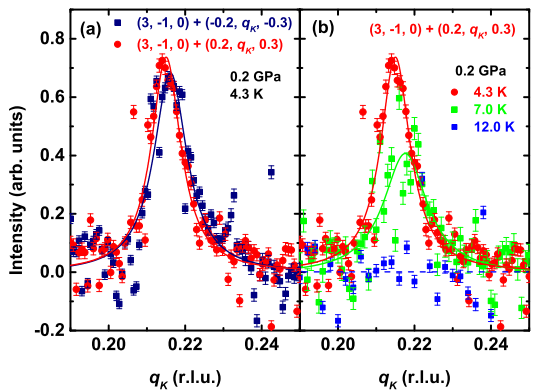}
\caption{(a) The projected \bm{$ \mathrm{Q1} $} and \bm{$ \mathrm{Q2} $} CDW peak profiles along \bm{$\mathrm{K}$} near (3, -1, 0) at $ T $ = 4.3 K and $ P $ = 0.2 GPa. (b) The $T$-dependence of the projected \bm{$ \mathrm{Q2} $} CDW peak profile at $ P $ = 0.2 GPa. A linear background is subtracted from all the data. Solid lines are Lorentzian fits to the data. The dashed line is a guide to the eye. Error bars correspond to 1 SD.
\label{Fig3}}
\end{figure}
%

\begin{figure}
\centering
\includegraphics[width=8.5cm]{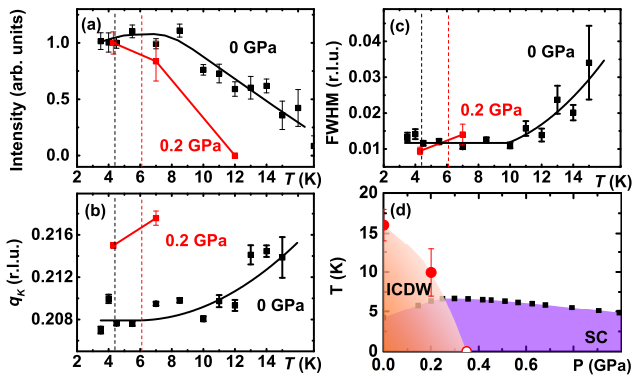}
\caption{The $T$-dependence of (a) the normalized peak intensity, (b) the peak position, and (c) the full-width at half-maximum (FWHM) along \bm{$\mathrm{q_{k}}$} for the \bm{$ \mathrm{Q1} $} CDW at 0 GPa and \bm{$ \mathrm{Q2} $} CDW at 0.2 GPa. The peak intensity shown in (a) is normalized to its value at 4.5 K for ambient $P$ and its value at 4.3 K for 0.2 GPa. Solid lines guide the eye. Black and red dashed lines indicate $ T_{c} $ at ambient $P$ and 0.2 GPa \cite{Jiao2014,Jo2017}. (d) $T$-$P$ phase diagram of \tpt. Empty red symbol at 0.35 GPa is a null data point, where no CDW peak is seen \cite{Supplemental}. The phase boundary of the ICDW phase, determined by the two data points at 0 and 0.2 GPa as well as the null data point at 0.35 GPa, is also indicated in the Fig. \ref{Fig:CDW_SC_plot} by the red dashed line. $ T_{c} $ values are from Ref. \cite{Pan2015}. 
\label{Fig4}}
\end{figure}

To confirm our observations at ambient $P$, and to further explore the $T$-$P$ phase diagram, we also conducted X-ray diffraction under pressure at a different synchrotron source \cite{Supplemental}. With the lowest pressure achieved in our measurement ($ P $ = 0.2 GPa \cite{Supplemental}), we again observed the \bm{$ \mathrm{Q1} $} and \bm{$ \mathrm{Q2} $} CDWs near the (3 -1 0) Bragg peak, at $ T $ = 4.3 K, as shown in Fig.~\ref{Fig3}(a). The ratio of the integrated intensity of the CDW peak and the nearby Bragg peak is again $\sim 10^{-3} $, in good agreement with our ambient-$P$ results. The intensities of the two CDWs are comparable within error, consistent with the picture of one checkerboard CDW. The $T$-dependence of the \bm{$\mathrm{q_{k}}$} line cuts are measured for the \bm{$ \mathrm{Q2} $} CDW at $ P $ = 0.2 GPa. The results as seen in Fig.~\ref{Fig3}(b) clearly show that the peak intensity is quickly suppressed with increasing $T$, and becomes undetectable at 12 K. Moreover, as shown in Fig. S3, our experiments do not show any sign of CDW \cite{Supplemental} at higher $P$ (0.35 GPa), where the peak position of the reported SC dome resides \cite{Pan2015,Jo2017}. Therefore, we find that the CDW becomes weaker, characterized by a smaller $ T_{\textrm{CDW}} $, with increased $P$, and disappear at higher $P$ when the SC is strongest (maximum $ T_{c} $). This seems to suggest a competitive nature of the interplay between CDW and SC, and is consistent with the fact that the ambient-$P$ CDW peak intensity plateaus at 9 K $\sim$ 5 K, below which the SC emerges and CDW peak intensity is weakened [see Fig. \ref{Fig2}(c) and Fig. \ref{Fig4}(a)]. However, as discussed below, we do not rule out the effect of quantum fluctuations near a CDW QCP on the SC. In cuprates, for example, while a reduced CDW intensity below $ T_{c} $ was reported as evidence for a CDW competing with SC \cite{Chang2012b}, a recent study also revealed the importance of a charge order quantum critical point \cite{Lee2020}.

The $T$-dependence of the CDW peak position along \bm{$\mathrm{q_{k}}$}, as shown in Fig.\ref{Fig2}(c) and Fig.\ref{Fig4}(b), shows some unexpected behaviors. At ambient $P$ the CDW peak center along \bm{$\mathrm{q_{k}}$} shifts towards a commensurate value 0.2 (a real-space periodicity of 5) with decreasing $T$, but saturates before locking into that value. The peak also shifts further away from $ q_{k} = 0.2 $ under pressure. Therefore, it appears that the CDW remains incommensurate along \bm{$\mathrm{q_{k}}$} down to the lowest $T$. Typically, one expects the CDW wavevector to evolve continuously with $T$ until abruptly jumping to a commensurate value, in a first-order lock-in phase transition \cite{Feng2015}. Examples include Q1D TTF-TCNQ and Q2D TaSe$_2$. However, many other materials show an evolution of the CDW wavevector without an eventual lock-in \cite{Monceau2012}, such as Q1D NbSe$_3$ and TbTe$_3$ and Q2D NbSe$_2$. Our observations indicate that \tpt\ falls into this latter category.

The CDW wavevectors, \bm{$ \mathrm{Q1} $} = [-0.2 0.21 -0.3] and \bm{$ \mathrm{Q2} $} = [0.2 0.21 0.3], featuring modulations along all three axes, appear at first to be at odds with the CDW reported in the previous STM study, which found a period-4 commensurate charge modulation only along the \bm{$ \mathrm{b} $}-axis \cite{Fan2015}. This discrepancy underlies the importance of the synchrotron X-ray measurement in directly observing the bulk CDW, in contrast to the STM, which is sensitive only to the projection of the charge modulation onto the sample surface [i.e. the (-1 0 3) cleavage plane]. A careful analysis of the \bm{$ \mathrm{Q1} $} and \bm{$ \mathrm{Q2} $} wavevectors and their projections on different planes (see Fig. S2 and the associated discussions in Supplemental Material \cite{Supplemental}) confirms that the Q1 and Q2 wavevectors revealed in our X-ray data actually reconciles the apparently conflicting results from STM \cite{Fan2015} and transport measurements \cite{Helm2017}.

The CDW correlation length can be extracted from the width of the CDW peak, which is measured by the full width at half maximum (FWHM) of the Lorentzian fits [Fig. 1(b) and Fig. 2, also see Supplemental Material \cite{Supplemental} for more details]. The $T$-dependent FWHM at $ P $ = 0 and 0.2 GPa are plotted in Fig.~\ref{Fig4}(c). At low $T$, FWHM $\sim$ 0.01 r.l.u., suggesting a CDW correlation length $ \xi_{CDW} $ $ \sim $ 120 \AA\ along \bm{$\mathrm{q_{k}}$} \cite{Supplemental}. A slight increase of FWHM to 0.015 r.l.u. is seen below $ T_{c} $, suggesting a possible competitive interplay of CDW with SC. Near $ T_{\textrm{CDW}} $ the FWHM increases drastically, and the inferred $ \xi_{CDW} $ reduces to $ \sim $ 30 \AA, which is close to the CDW wavelength (i.e. CDW modulation period) along the \bm{$ \mathrm{b} $}-axis $ \lambda_{b} $ $\sim$ 5$b$ $\sim$ 20 \AA, signaling the melting of the static CDW at higher $T$.

%
%

In Fig.~\ref{Fig4}(d) we establish the $T$-$P$ phase diagram using $ T_{\textrm{CDW}}(P) $ from our X-ray diffraction measurements and $ T_{c} $ from Ref.~\cite{Pan2015}. The SC dome peaks at the pressure where the CDW vanishes. Although such observations may generally be expected within standard BCS theory (i.e. $ T_{c} $ increases as the density of states released upon CDW gap closure becomes available for the formation of Cooper pairs), we note that other scenarios including SC related to a CDW QCP \cite{Gruner2017} may not be ruled out: (a) the energy scales of the CDW and SC are comparable, while the former is typically much larger than the latter in conventional superconductors; and (b) the CDW transition in \tpt\ appears to be a second order phase transition, as hinted by the very weak kink-like feature in resistivity \cite{Helm2017} and the smooth evolution of the CDW order parameter seen in our X-ray data, although we do not have evidence regarding whether the transition remains second order or sufficiently weak first order at $T=0$.

Our results reveal a delicate checkerboard CDW in \tpt\ (see the Supplemental Material \cite{Supplemental}). It is found that \tpt\ features (a) a character of mixed dimensionality, (b) two CDWs related by a lattice symmetry developing at the same $T$, (c) CDW wavevectors that are commensurate along two lattice directions but incommensurate along the third even at low $T$, (d) no clear signatures of FS reconstruction as the CDW develops (see our dHvA oscillation results), and (e) an interesting interplay of the CDW and SC. The uniqueness of the CDW in \tpt\ is further demonstrated when comparing to other CDW materials.

On the one hand, \tpt\ is often considered Q1D, structurally similar to the transition metal trichalcogenides such as NbSe$_3$, NbS$_3$, ZrSe$_3$, and TaSe$_3$, each of which has a monoclinic phase formed from chains within planes, and also the rare-earth tri-tellurides $R$Te$_3$ ($R=$Sm, Gd, Tb, Dy, Ho, Er, Tm) with weakly orthorhombic structures. All these materials host multiple CDWs~\cite{Monceau2012}, which, unlike \tpt, are not related by a lattice symmetry, and develop at different $T$ (also see Supplemental Material \cite{Supplemental}). On the other hand, the resistivity ratios of \tpt\ suggest that the material is better thought of not as Q1D but as a dimensional hybrid with Q1D chains within Q2D planes within a 3D material. DFT calculations confirm the existence of FS pockets of each dimensionality~\cite{Singh2014}. Thus a fairer comparison might be found with Q2D CDW materials such as the transition metal dichalcogenides (TMDC) $2H$-NbSe$_2$, $2H$-TaSe$_2$, and $1T$-TaS$_2$~\cite{Giambiattista1990,Feng2015}, which, however, feature CDW entirely within the Q2D planes. 

%
\begin{figure}
\centering
\includegraphics[width=7.5cm]{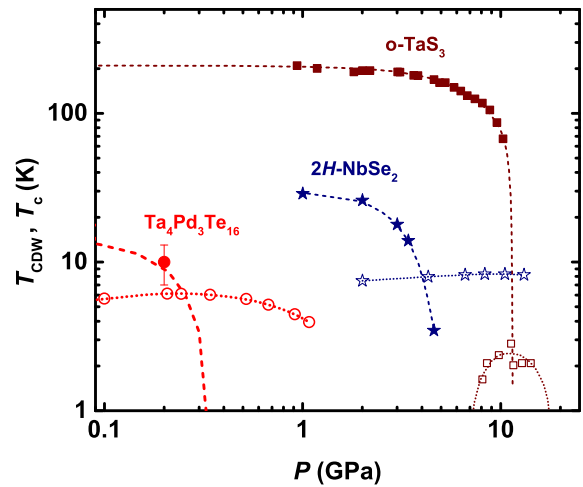}
\caption{
Phase diagram of a few selected compounds, as a function of $T$ and $P$, showing $T_\textrm{CDW}$ (solid symbols) and $T_c$ (empty symbols). For \tpt, the CDW phase boundary, determined from our X-ray data at 0, 0.2, and 0.35 GPa and indicated by the red dashed line, is shown along with $T_c$ values from Ref.~\cite{Jo2017}. Note that logarithmic scales are used for both axes due to the vastly different scales for the different materials. As a result, our data points at $P$ = 0 and $T_\textrm{CDW}$ $\sim$ 0 become invisible, though the red dashed line remains as a good indicator for the CDW phase boundary. Other compounds: $2H$-NbSe$_2$ \cite{Feng2015,Leroux2015}; o-TaS$_3$ \cite{NunezRegueiro1992,NunezRegueiro1993}.
\label{Fig:CDW_SC_plot}}
\end{figure}

It has previously been argued that CDWs never truly develop by a Q1D Peierls-type mechanism~\cite{Johannes2008,Zhu2015,Flicker2015}. From this point of view, all compounds with a purported low-dimensional character might more reasonably be thought of as dimensional hybrids, with \tpt\ simply being a particularly clear example. A consequence can be seen in the behavior of the CDW under pressure, and its interaction with SC. As noted in Ref.~\cite{Monceau2012}, increasing $P$ has the effect of increasing interchain (or interplane) coupling, which might have one of two opposite effects on $T_\textrm{CDW}$ depending on how Q1D the system really is. In extreme Q1D systems, $T_\textrm{CDW}$ is heavily suppressed by fluctuations to well below the value predicted by mean-field theory~\cite{Lee1973}. The reduction in interchain spacing (increased interchain coupling) decreases the one-dimensional character of the system, suppressing these fluctuations, and increasing $T_\textrm{CDW}$. On the other hand, if $T_\textrm{CDW}$ is not dictated by Q1D fluctuations, the primary effect of the increased interchain coupling is to decrease the density of states at the Fermi level, suppressing $T_\textrm{CDW}$. In the standard BCS picture, the phonon mode formerly suppressed by the CDW could become available to mediate SC at high $P$ and an SC phase could develop, although the resulting SC phase does not necessarily form a dome centered around the vanishing of the CDW. As shown in Fig. \ref{Fig:CDW_SC_plot} and Fig. S4, such a picture seems to apply to many Q2D and 3D materials (e.g. $2H$-NbSe$_2$ \cite{Leroux2015}, $2H$-TaS$_2$ \cite{Grasset2019,Freitas2016}, $1T$-TaS$_2$ \cite{Sipos2008}, $1T$-TaSe$_2$ \cite{Wang2017}, $1T$-TiSe$_2$ \cite{Kusmartseva2009}, (Ca,Sr)$_3$Ir$_4$Sn$_{13}$ \cite{Klintberg2012,Goh2011}) (see Supplemental Material \cite{Supplemental}), as well as to \tpt\ to a certain extent. However, as explained above, we do not rule out the enhancement of $T_c$ due to possible QCP-related fluctuations in \tpt\ and the possibility of an incommensurate density wave quantum criticality in 2D metals has recently been discussed \cite{Halbinger2019}. 

In summary, we report a direct observation of a bulk CDW with its checkerboard wavevectors in \tpt. It shows several interesting characteristics: (a) a unique mixed character of dimensionality and (b) the lowest value of the $T_\textrm{CDW}$/$T_c$ among prototypical CDW materials. Future studies with probes sensitive to both CDW and SC (e.g. High-$P$ Raman spectroscopy, ARPES, Second Harmonic Generation, etc) are desirable in order to fully understand this interesting interplay.

\begin{acknowledgments}

We acknowledge useful discussions with S. H. Simon, and experimental help from J. Strempfer. We are thankful to J. Analytis for supervising the sample synthesis efforts, and to the Analytis Lab for contributing facilities for sample synthesis and single crystal growth. We thank Roger Sommer for help with single crystal X-ray diffraction at the METRIC (Molecular Education, Technology, and Research Innovation Center) at North Carolina State University. We thank Sven Friedemann for providing unpublished data on 2$H$-NbSe$_{2}$ which is shown in the Supplemental Material Fig. S5. This work is based upon research conducted at the Cornell High Energy Synchrotron Source (CHESS) which is supported by the National Science Foundation under award DMR-1332208 and DMR-1829070. Work at Argonne was supported by the U.S. DOE Office of Science, Office of Basic Energy Sciences, under Award No. DE-AC02-06CH11357. F.F. acknowledges support from the Astor Junior Research Fellowship of New College, Oxford. Z.S., W.M.S., S.D., and S.H. acknowledge support provided by funding from the Powe Junior Faculty Enhancement Award, and William M. Fairbank Chair in Physics at Duke University. A portion of this work was performed at the National High Magnetic Field Laboratory, which is supported by National Science Foundation Cooperative Agreement No. DMR-1157490 and the State of Florida.\\

S.H. conceived and supervised research; T.H. grew the sample; S.K., J.L., J.R., D.H., and S.H.
conducted X-ray diffraction study at ambient $P$; Z.S., W.S., S.D., G.F., D.H., and S.H.
conducted X-ray diffraction study at high $P$; Z.S., D.G., and S.H. conducted dHvA
oscillation measurements; Z.S., F.F., and S.H. wrote the paper.\\

Z.~S.~and S.~K.~contributed equally to this work.\\

\end{acknowledgments}


\end{document}